\documentclass[prl,showpacs,twocolumn,superscriptaddress]{revtex4}
\usepackage{amsmath}
\usepackage{graphicx} % Include figure files
\usepackage{dcolumn}  % Align table columns on decimal point
\usepackage{bm}       % bold math


\begin{document}

\author{Stephanie Manz}
 %\email{manz@physi.uni-heidelberg.de}
 \affiliation{Physikalisches Institut, Universit\"at Heidelberg, 69120 Heidelberg, Germany}
 \affiliation{Atominstitut \"Osterreichischer Universit\"aten, TU-Wien, A-1020 Vienna, Austria}
\author{Thomas Fernholz}
 %\email{fernholz@physi.uni-heidelberg.de}
 \affiliation{Physikalisches Institut, Universit\"at Heidelberg, 69120 Heidelberg, Germany}
 \affiliation{Van der Waals-Zeeman Institute, University of Amsterdam, 1018 XE Amsterdam, The Netherlands}
\author{J\"org Schmiedmayer}
 %\email{schmiedmayer@atomchip.org}
 \affiliation{Physikalisches Institut, Universit\"at Heidelberg, 69120 Heidelberg, Germany}
 \affiliation{Atominstitut \"Osterreichischer Universit\"aten, TU-Wien, A-1020 Vienna, Austria}
\author{Jian-Wei Pan}
 %\email{pan@physi.uni-heidelberg.de}
 \affiliation{Physikalisches Institut, Universit\"at Heidelberg, 69120 Heidelberg, Germany}
\title{Collisional decoherence during writing and reading quantum states}
\begin{abstract}

Collisions, even though they do not limit the lifetime of quantum
information stored in ground state hyperfine coherences, they may
severely limit the fidelity for quantum memory when they happen
during the write and read process.  This imposes restrictions on
the implementation of Raman type quantum processes in thermal
vapor cells and their performance as a quantum memory. We study
the effect of these collisions in our experiment.
\end{abstract}
\date{\today}
\maketitle

 \hyphenation{inter-fero-me-ter}

%% -----------------------------------------------------------------------------
%%
%% Introduction -> Jian-Wei
%%
%% -----------------------------------------------------------------------------

In recent years, significant experimental advances have been
achieved in the field of quantum communication
(QC)~\cite{Gisin2002,Zeilinger2005}. However, photon loss and
detector noise limit direct QC to moderate distances (up to 100 km
in quantum cryptography).  In 2001, Duan, Lukin, Cirac and Zoller
(DLCZ) proposed a practical quantum repeater \cite{DLCZ,Briegel98}
based on writing and reading single excitations in atomic
ensembles using Raman type processes. A joint projective
measurement of individual photons emitted from two separated
atomic ensembles leads to qubit-type entanglement of collective
excitations in both ensembles, which combined with entanglement
swapping~\cite{Zukowski1993} and entanglement
purification~\cite{Bennett1996} allows to create entanglement over
(arbitrary) long distances. Essential to the scalability of the
DLCZ scheme are the long lived collective excitations, which
represent a quantum memory. Without such quantum memory, the
overhead scales exponentially with the channel length.

Although entanglement swapping~\cite{Pan1998} and entanglement
purification~\cite{Pan2003} have been experimentally demonstrated
with linear optics, it is difficult to achieve the high fidelity
quantum memory. Significant experimental advances have been made
to implement the DLCZ--scheme \cite{DLCZ} both with ultra-cold
atomic clouds
\cite{Kuzmich03,chou05,chaneliere05,Matsukevich04,Kuzmich2006single,Shuai06}
and hot vapors in buffer gas cells
\cite{Lukin2003,andre05,Eisaman05}. While ultra-cold ensembles
require substantial technological effort, vapor cells provide
comparatively easy experimental access. Moreover atomic clock
experiments have shown that with the correct coating of cell walls
and the use of buffer gas the ground state coherence can be
preserved for up to $> 10^{8}$ collisions \cite{happer74} leading
to very narrow line widths.

In this paper, we demonstrate however that the mapping process
between light fields and atomic coherence is strongly influenced
by collisions.  This severely limits the fidelity of the
DLCZ-scheme and a quantum memory in hot atomic ensembles.

The mapping process between light fields and atomic coherence is
essential to the DLCZ scheme \cite{DLCZ}.  Single excitations are
written in an atomic ensemble Raman transitions in a three level
$\lambda$-configuration: (ground states $|1\rangle$, $|2\rangle$,
excited state $|3\rangle$). First, the atoms are optically pumped
to $|1\rangle$. Then a write--laser, detuned by $\Delta_{write}$
from the $|1\rangle$$\rightarrow$$|3\rangle$ transition induces a
spontaneous Raman process. Upon detection of a \emph{Stokes}
photon the atoms are projected to a long--lived collective state
(the quantum memory), corresponding to a spin-wave excitation.
This process can be inverted with a read laser detuned
($\Delta_{read}$) from the $|2\rangle$$\rightarrow$$|3\rangle$
transition, which converts the spin excitation back to
\emph{Anti--Stokes} photons.

%% -----------------------------------------------------------------------------
%%
%% Our setup
%%
%% -----------------------------------------------------------------------------

\begin{figure}[t]
\begin{center}
\includegraphics[width=\columnwidth]{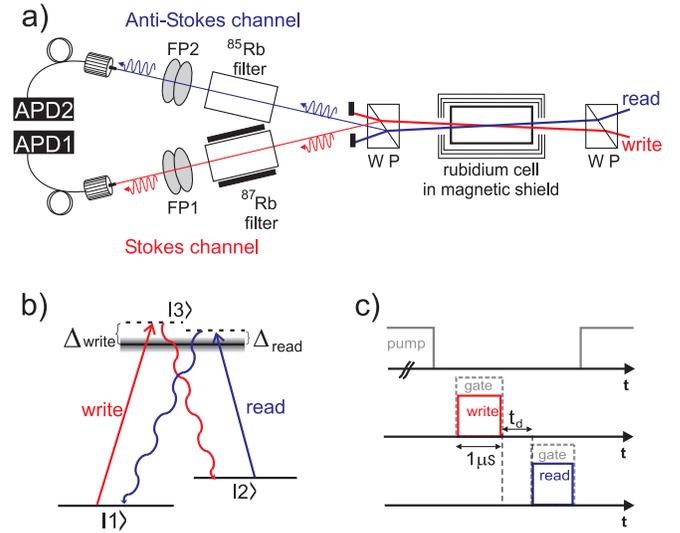}
\caption{(color online) \textit{a)} Experimental setup: Two
control lasers (write and read) are combined at a Wollastone prism
(WP) and intersect inside a magnetically shielded $^{87}$Rb vapor
cell with Neon buffer gas. The weak signal fields (Stokes and
Anti-Stokes) are separated from the control beams at a second
polarizer. After spectral filtering, the signal photons are
detected by fiber coupled avanlanche photo diode (APD).
\textit{b)} Simplified level scheme: A lambda system is used to
map ground state coherences to the signal fields via two detuned
control beams. \textit{c)} Laser pulse timing for spontaneous
creation of ground state coherence and delayed read-out.}
\label{fig:setup}
\end{center}
\end{figure}

In our experimental implementation (Fig.~\ref{fig:setup}) we use
hot $^{87}$Rb vapor as the atomic medium.  The levels forming the
$\lambda$-system are given by the $5^2S_{1/2}, F=1$ and
$5^2S_{1/2}, F=2$ ground states (corresponding to $|1\rangle$ and
$|2\rangle$), and the $5^2P_{1/2}$ manifold comprised in the
excited state $|3\rangle$. The \emph{write} and \emph{read}
processes are controlled by two nearly co-propagating laser beams
with orthogonal linear polarizations ($P_{write}\approx 1\textrm{
mW}$ and $P_{read}\approx 5\textrm{ mW}$) focussed to $\approx
300\textrm{ $\mu$m}$ diameter intersecting inside the cell at a
small angle $\alpha\approx 5\textrm{ mrad}$.  For linear
polarizations, the signal beams are mainly generated with
polarizations orthogonal to their respective control
beams~\cite{weis}, and can be separated at a crystal polarizer
(Wollaston prism). The signals are detected with electronically
gated, fibre coupled avalanche photon detectors (PerkinElmer,
SPCM-AQR series). This set--up resembles the one desribed
in~\cite{Lukin2003}.

To obtain high enough optical depth, the Rb is heated to
T=60$^{\circ}$C, and $p_{Ne}\approx7\textrm{ torr}$ of Neon buffer
gas keeps the hot atoms long enough in the interaction region. All
is confined in a 4 cm long glass-cell inside a three layer
$\mu$-metal shield to reduce magnetic stray fields.

In order to achieve a high enough signal-to-noise ratio, we employ
various filter techniques to suppress leakage from the strong
\textit{write} (\textit{read}) beams. Before entering the atomic
ensemble FP filter cavities reduce the spectrally broad amplified
spontaneous emission (ASE) from the \textit{write} (\textit{read})
lasers. These resonators also provide spatially clean beam
profiles. Due to the limited extinction ratio of the crystal
polarizer ($\approx 10^{-5}$), the strong control beams leak into
the respective signal channels, and further filtering is
essential. The \textit{write} (\textit{read}) lasers are blue
detuned from the $F=1(2)\rightarrow F'=2$ transitions, $\Delta$
ranging between $0.5\textrm{ GHz}$ and $1.5\textrm{ GHz}$. The
\emph{read} channel is filtered by a hot, isotopically enriched
$^{85}$Rb vapor cell, which is opaque for the control
(\emph{read}) but transparent for the signal
(\textit{Anti-Stokes}) frequency. For the \textit{write} channel,
we use an optically pumped $^{87}$Rb vapor cell that blocks
residual \emph{write} light in the \textit{Stokes} signal. The
absorption of the filter cell is broadened by a magnetic field of
$\approx 0.1\textrm{ Tesla}$ using permanent magnets, allowing for
larger detunings of the \textit{write} laser.

%% -----------------------------------------------------------------------------
%%
%% Frequency resolving scans
%%
%% -----------------------------------------------------------------------------
%
\begin{figure}[t]
\begin{center}
\includegraphics[width=8.5cm]{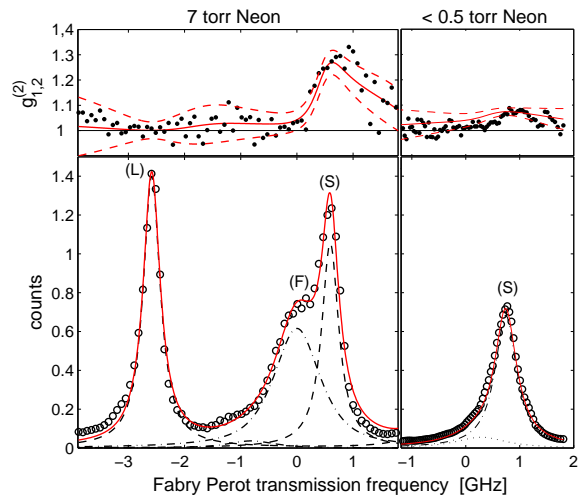}
%                                                         %
%Data:2005-03-16\ohne747576.dat and 2005-03-19\scan192.dat%
%                                                         %
\caption{(color online)(\textit{left}) Frequency resolved scan of
the Stokes channel with imperfectly filtered \emph{write} light as
a frequency reference. The length of a Fabry-Perot interferometer
(FP1, FSR=5 GHz) was scanned while the signal intensity (bottom)
and its cross correlation to the Anti-Stokes channel (top) were
recorded. The model (solid red line) includes the coherent (S) and
incoherent (F) contributions to the \emph{Stokes} signal, and the
\emph{write} laser light (L) that appears again two FSR away from
its original frequency. The data (black dots) show that only the
coherent Stokes contribution is correlated to the Anti-Stokes
signal. Statistical uncertainties in the measured cross
correlation  were calculated according to the finite number of
experimental cycles. The red dashed lines indicate confidence
bands for one sigma deviation. (\textit{right}) Corresponding
plots for the mean of four measurements with reduced buffer gas
pressure below 0.5 torr.} \label{fig:collevalu}
\end{center}
\end{figure}
In order to further examine the generated quantum fields in
detail, we employ additional Fabry-Perot etalons (free spectral
range FSR=5 GHz with FWHM-linewidth of 480 MHz), which allow us to
perform frequency resolving scans of the signal channels.
Fig.~\ref{fig:collevalu} shows a scan of the Stokes channel, while
the Fabry-Perot interferometer in the Anti-Stokes channel was
tuned to maximum transmission (0.8 counts per pulse). Each data
point represents 4000 experimental cycles, which were performed at
a repetition rate of 10 kHz. The bottom left of
Fig.~\ref{fig:collevalu} shows the intensity spectrum, represented
by the mean photon number per pulse. As a frequency reference, we
included a small fraction of the \emph{write} light by reducing
the temperature of the filter cell, labelled as (L) in
fig.~\ref{fig:collevalu}.  The \emph{write} light was blue detuned
from the $F$=1$\rightarrow F'$=2 transition by 800 MHz. We find
\textit{Stokes} light at a relative red detuning corresponding to
the hyperfine splitting frequency of $^{87}$Rb (6.83 GHz). This
frequency is amplified by the medium, and the intensity of this
signal can be controlled by the length of the \emph{write} pulse
and its detuning from the D1 transition frequency.

The striking observation is that for a vapor cell with buffer gas
($\sim 7$ torr Neon), an additional contribution to the signal is
present, close to the Stokes frequency. The intensity of this
signal is independent of the pulse length, and it only becomes
significant at the single (few) photon level, where amplification
of the Stokes light is negligible. Scans of the retrieve channel
revealed a similar signal close to the \textit{Anti-Stokes}
frequency.

\begin{figure}[tb]
\begin{center}
\includegraphics[width=8.5cm]{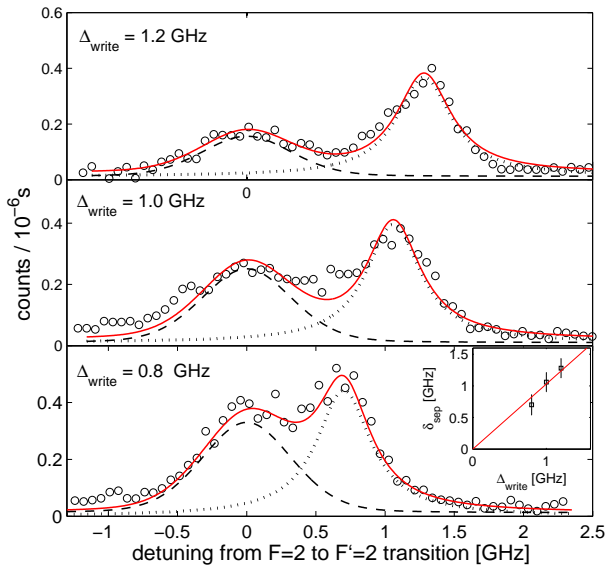}
\caption{(color online) Coherent and incoherent contributions to
the Stokes signal for various \emph{write} detunings. The three
upper plots show frequency scans of the Stokes channel, using a
Fabry-Perot resonator with a linewidth of 480 MHz and an FSR of 5
GHz. The solid line is a fitted model, including a sharp Stokes
peak (dotted) and a broader fluorescence peak (dashed) convoluted
with the Fabry-Perot transmission function. The width of the
fluorescence is assumed to be given by the Doppler profile
according to the atomic temperature. The inset displays the
approximate equality of peak separation and write detuning.}
\label{fig:all_blue}
\end{center}
\end{figure}

We attribute the additional signal to incoherent Raman scattering
of the \emph{write}(\emph{read}) light, caused by collisional
perturbation of the excited state during the \textit{write}
(\textit{read}) process. This process is included in mechanisms
described in~\cite{Rousseau} and is similar to collision induced
fluorescence reported in~\cite{Raymer1977} for the strongly
stimulated regime. Its frequency should be resonant with the
atomic transition. For a more quantitative comparison, we model
the incoherent signal using two Voigt lines with a separation of
812 MHz, given by the hyperfine splitting of the excited state.
The FWHM-line widths are fixed to 480 MHz for the Gaussian part,
given by the Doppler broadening of Rubidium at a temperature of
$60^{\circ}$C, and the Fabry-Perot line width for the Lorentzian
part. We find a significant contribution from the upper excited
level only. To support this interpretation we performed similar
measurements at different detunings (Fig.~\ref{fig:all_blue}). The
frequency difference between the two components equals the
\emph{write} detuning from the $F=2\rightarrow F'=2$ transition.

With the frequency resolved detection we can investigate the
coherent nature of the written collective excitation. The upper
part of Fig.~\ref{fig:collevalu} shows the measured cross
correlation $g^{(2)}_{1,2}=\langle n_{1}n_{2}\rangle/\langle
n_{1}\rangle\langle n_{2}\rangle$ between the two signal beams.
The solid line gives the prediction under the assumption that
collisions during the write process lead to a completely dephased
excitation.

While the observed cross correlation is independent of the
detection efficiencies, it depends on the signal-to-noise ratios
$\chi_{1}$ and $\chi_{2}$ in the two channels as $g^{(2)}_{1,2}-1
= (g^{(2)}_{S,AS}-1)(1+1/\chi_{1})^{-1}(1+1/\chi_{2})^{-1}$, where
$g^{(2)}_{S,AS}$ is the cross correlation between the noise-free
signals. Our data can be described by a single free parameter
$(g^{(2)}_{S,AS}-1)(1+1/\chi_{2})^{-1}=0.35$. The good agreement
of our data with this model shows that only upon detection of
\textit{coherently} scattered Stokes light, the ground state
coherence can be converted into correlated Anti-Stokes light
during the read pulse, and collisions during the read and write
pulse lead to destruction of the quantum excitation.

To support our assumptions, frequency resolving scans have also
been taken with strongly reduced buffer gas pressure ($\sim0.2$
torr). In this experiment the generation of coherent Stokes light
was enhanced, and we lowered the temperature of the Rb vapor (Rb
density) to obtain similar count rates in the Stokes channel for
the same \emph{write} parameters. At low buffer gas pressure, no
significant incoherent contributions are apparent. However, the
detection of cross correlation was hampered by the fast atomic
diffusion. Atoms leave the interaction region too quickly,
resulting in a strongly reduced retrieve efficiency and a low
signal-to-noise ratio in the \textit{Anti-Stokes} channel.

%% -----------------------------------------------------------------------------
%%
%% Theoretical effects of homogeneous broadening
%%
%% -----------------------------------------------------------------------------
%
The observed decoherence processes during write and read might be
related to a recent theoretical work  by Childress et
al.~\cite{childress05}.   They predict the appearance of
incoherently scattered light for color centers in diamond. They
model the perturbations of the excited level as a time dependent
detuning with white noise characteristics under the assumption
that the perturbing mechanism is slow compared to the optical
frequency but faster than the radiative lifetime of the excited
state. Their model predicts coherent and incoherent signal
contributions with a relative frequency difference equal to the
detuning of the driving laser from the atomic transition.  The
relative weights are given by $\gamma:\Gamma$, where $\Gamma$ is
the fluctuation amplitude of the excited state detuning and
$\gamma$ its radiative decay rate.

The model qualitatively also describes our findings, especially
the frequency difference between the coherent and incoherent part
of the spectrum (Fig.~\ref{fig:all_blue}). In our case the
perturbations are given by the collisions, which behave like delta
function perturbations. From a broadening
coefficient~\cite{Bemerkung} of $\gamma_c \approx (7\pm 2)
MHz/torr$~\cite{Izotova} we estimate a collision rate for atoms
undergoing a Raman transition of $\gamma_c \approx 49 MHz $ at 7
torr Neon gas pressure. For these parameters collisions happen on
a similar timescale compared to the D1 decay rate of $\gamma=36
\times 10^{-6}s^{-1}$. For very low \emph{Stokes} amplification we
observe the relative weights between the coherent and incoherent
part to be roughly equal. In the strong collision limit their
ratio should be given by the branching ratio $\gamma_R \approx
\gamma_c:\gamma$ of excited state decay with and without
intermediate collisions.  An accurate description would require an
elaborate model that includes a detailed description of the
coherence in the Raman process in this collisional regime,
inhomogeneous doppler broadening and velocity redistribution
induced by the collisions, effects of collective enhancement in
the atomic ensemble and the substructure of the excited level.
%
%%-----------------------------------------------------------------------------
%%
%%
%% -----------------------------------------------------------------------------
%%
%% Conclusions
%%
%% -----------------------------------------------------------------------------
%
%
Our experiments have shown, that disturbances during writing and
reading collective excitations are very important for the
performance of quantum processes. Whereas one usually selects very
stable protected states to store the quantum information, the
states involved during reading and writing can be very fragile. If
one uses the hyperfine ground states of alkali atoms, the
coherence in the nuclear spin states is shielded from collisions
by the electrons of the atom, as given by dark state coherence or
under EIT conditions. Using special coated cells and noble buffer
gas the atoms can endure many ($>10^8$) collisions before loosing
hyperfine coherence. The Raman transition based write and read
processes of a quantum memory involves electronically excited
states, and a single collision can destroy the coherence, and
(quantum) information is lost.

These write and read errors can be very severe in quantum
applications. For some, like the quantum repeater they can be
filtered out by selecting only photons on the unperturbed Raman
line. But even then they lead to increased overhead, since the
fidelity of a single step is reduced. To keep the polynomial
scaling one must make sure that the overhead does not lead to
exhaustion of the memory coherence time. For a quantum memory the
decoherence in the write and read processes are more severe and
directly reduce fidelity.

Consequently, to achieve high fidelity quantum operations these
incoherent processes need to be suppressed. For thermal ensembles
this requires reducing the probability of collisions during the
read and write process, which means reducing the buffer gas
pressure.  This leads to a greatly increased diffusion length. To
keep a good retrieve efficiency and long storage time one must
keep the atoms in a tight overlap with read and write beams. For
an increased diffusion length this can be done by a larger beam
diameter. This makes the filtering of unwanted light much harder,
since the required total power has to be increased. A better way
might be to put the atoms in a hollow fiber~\cite{benabid05},
which can also be used to guide the read and write light. The
atoms are then kept in the laser beam by collisions with the
walls, and one can refrain from using buffer gas at all. For a
thermal velocity of $\sim 300 \; m/s$ the typical atom wall
collision rate will be $\sim 3 \times 10^6$ in a 100 $\mu m$
diameter hole. Again collisions between the wall and atoms in the
excited state will lead to decoherence, and should be avoided. One
can achieve this by forcing the light field to be zero at the
wall. Working far off resonance, the atoms will evolve
adiabatically to the ground state when being excited by the light
field in approaching the wall. This would require the fiber hole
to be coated by a mirror which also protects the stored atomic
spin. A different approach could be to design the guided light
field by a photonic band gap structure, keeping it minimal close
to the walls.

In conclusion we point out that collisions, even though they might
not limit the life time of a stored ground state hyperfine
coherence, dark states or EIT, they may severely limit the
fidelity for this type of quantum memory when they happen during
the write and read process. These collisions, be it with the
buffer gas atoms or cell walls are required to keep the atoms in
the write / read volume when using thermal vapor cells. Cold atoms
have a great advantage because they allow wall free confinement,
and the collision processes happen at timescales many orders of
magnitude slower than the read and write processes.

This work was supported by the Deutsche Forschungsgemeinschaft E.
Noether programm contract number PA938 and European Union,
Integrated Project FET/QIPC "SCALA". We thank Igor Mazets for
helpful discussions.


\begin{thebibliography}{16}
\bibitem{Bemerkung} We have to consider the collisional broadening
and not the ground--state collision rate. They differ due to the
different electrical polarizability for ground and excited states.
We could not directly measure the collisional broadening in the
measurement cell at the operating temperature.
\end{thebibliography}
\end{document}